\pdfoutput=1
\documentclass[12pt,aps,english]{article}
\usepackage{amssymb, amsmath}
\usepackage{multicol}
\usepackage{amsmath, amssymb}
\usepackage[mathscr]{eucal}
\usepackage{graphicx}
\usepackage{subfigure}
\usepackage{amsfonts}
\usepackage{babel}
\usepackage{framed}
\usepackage{color}
\usepackage[T1]{fontenc}
\newcommand{\bfi}{\bfseries\itshape}

\newcommand{\be}{\boldsymbol{\eta}}
\newcommand{\rhobar}{\overline{\varrho}}
\newcommand{\id}{{\mathrm{id}}}
\newcommand{\rem}[1]{}
\newcommand{\remfigure}[1]{}

\makeatletter
\@addtoreset{equation}{section}

\makeatother

\newtheorem{theorem}{Theorem}

\newtheorem{definition}[theorem]{Definition}

\newtheorem{lemma}[theorem]{Lemma}

\newtheorem{remark}[theorem]{Remark}

\numberwithin{theorem}{section}
\newenvironment{proof}[1][Proof]{\textbf{#1.} }{\ \rule{0.5em}{0.5em}}

\def\0{{\bf 0}}

\newcommand{\pp}[2]{\frac{\partial #1}{\partial #2}}

\begin{document}

\title{
Singular solutions of a modified 
two-component Camassa-Holm equation
}
\author{
Darryl D. Holm$^{1,\,2}$, Lennon \'O N\'araigh$^{3}$,  Cesare Tronci$^{1,4}\!$\\ \vspace{-.3cm}
\\
{\footnotesize $^1$ \it Department of Mathematics, Imperial College London,
180 Queen's Gate, London SW7 2AZ, UK}\\
{\footnotesize $^2$ \it Institute for Mathematical Sciences, Imperial College
London, 53 Prince's Gate, London SW7 2PG, UK
} \\
{\footnotesize $^3$ \it Department of Chemical Engineering, Imperial College London, London SW7 2AZ, UK}\\
{\footnotesize $^4$\,\it TERA Foundation for Oncological Hadrontherapy,
11 V. Puccini, Novara 28100, Italy}
\\ 
}

\date{\today}

\maketitle
\begin{abstract}\noindent
The Camassa-Holm equation (CH) is a well known integrable equation describing the velocity dynamics of shallow water waves. This equation exhibits spontaneous emergence of singular solutions (peakons) from smooth initial conditions.
The CH equation has been recently extended to a two-component integrable system (CH2), which includes both velocity and density variables in the dynamics. Although possessing peakon solutions in the velocity, the CH2 equation does not admit singular solutions in the density profile. 
We modify the CH2 system to allow dependence on average density as well as pointwise density. The modified CH2 system (MCH2) does admit peakon solutions in velocity and average density. We analytically identify the steepening mechanism that allows the singular solutions to emerge from smooth spatially-confined initial data. Numerical results for MCH2 are given and compared with the pure CH2 case. These numerics show that the modification in MCH2 to introduce average density has little short-time effect on the emergent dynamical properties. 
However, an analytical and numerical study of pairwise peakon interactions for MCH2 shows a new asymptotic feature. Namely, besides the expected soliton scattering behavior seen in overtaking and head-on peakon collisions, MCH2 also allows the phase shift of the peakon collision to diverge  in certain parameter regimes.
\end{abstract}




\section{Introduction}

Singular, measure-valued solutions in fluids appear in the familiar example of the point vortex solutions for the Euler vorticity equation on the plane. The point vortices are measure-valued solutions whose motion is governed by a type of multi-particle dynamics. In three dimensions this concept extends to vortex filaments or vortex sheets, for which the vorticity is supported on a lower dimensional submanifold (1D or 2D respectively) of the Euclidean space $\mathbb{R}^3$. These singular vortex solutions form an invariant manifold. That is, they persist under the evolution of the Euler inviscid fluid equations if they are present initially, but they are not created by Euler fluid motion from smooth initial conditions. Whether they may be created from smooth initial conditions by Navier-Stokes viscous fluid motion remains a famous open problem. 
Invariant manifolds of singular solutions also exist in plasma physics as magnetic field lines in magnetohydrodynamics and in Vlasov kinetic theory as single-particle trajectories. 

Shallow water theory introduces  in the limit of vanishing linear dispersion another class of singular solutions which has the property of emerging from smooth initial conditions. This class of {\bfi emergent singular solutions} is the main subject of the present paper. 

\subsection{Singular solutions for unidirectional shallow water waves}

The Korteweg-de Vries equation (KdV) for unidirectional shallow water waves is 
\begin{equation} \label{KdV-eqn1}
u_t + 3uu_x = -c_0u_x+\gamma{u}_{xxx}
\,,
\end{equation}
in which the fluid velocity $u$ is a function of time $t$ and position $x$ on the real line. The subscripts denote the corresponding partial derivatives, while the constants $c_0$ and $\gamma $ represent the effects of linear dispersion.
KdV appears at linear order in an asymptotic expansion for unidirectional shallow water waves on a free surface under gravity. The expansion is made in terms of two small dimensionless ratios for small-amplitude long waves in shallow water.  

At quadratic order in the same asymptotic expansion, the Camassa-Holm equation (CH) appears \cite{CaHo1993},
\begin{equation} \label{CH-eqn1}
m_t + um_x + 2 mu_x = -c_0u_x+\gamma{u}_{xxx}
\,,\qquad
m=u-\alpha^2u_{xx}
\,.
\end{equation}
In the limit that $\alpha^2\to0$, CH recovers KdV. 

Both KdV and CH are completely integrable bi-Hamiltonian equations describing the effects of nonlinearity and dispersion on unidirectional shallow water waves at their respective orders in asymptotic expansion. Moreover, both KdV and CH arise as compatibility conditions for their respective  {\bfi isospectral eigenvalue problem} and a linear evolution equation for the corresponding eigenfunctions. The properties of being bi-Hamiltonian and possessing an associated isospectral problem imply that the one-dimensional KdV and CH equations are each {\bfi completely integrable} as Hamiltonian systems. In particular, they each possess an infinity of conservation laws and each is solvable by its corresponding  {\bfi inverse scattering transform (IST)}. Perhaps not unexpectedly, the isospectral eigenvalue problem for CH differs from that for KdV but recovers the KdV case for $\alpha^2\to0$.
%
%
%
%
\begin{figure}[h!]
\begin{center}
\includegraphics[width=.75\textwidth,angle=0]{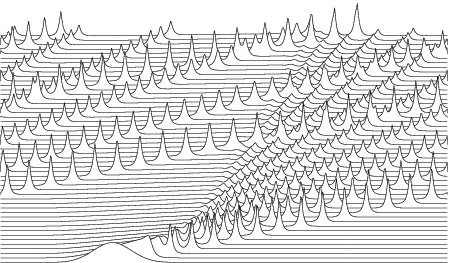}
\end{center}
\caption{\footnotesize
Under the evolution of the dispersionless CH equation (\ref{CH-eqn-intro}), an ordered {\bfi wave train of peakons} emerges from a smooth localized initial condition (a Gaussian). The spatial profiles at successive times are offset in the vertical to show the creation and  evolution of the peakons. The peakon wave train eventually wraps around the periodic domain, thereby allowing the leading peakons to overtake the slower peakons from behind in collisions that conserve momentum and preserve the peakon shape but cause phase shifts in the positions of the peaks, as discussed in \cite{CaHo1993}. This {\bfi soliton behavior} is the hallmark of completely integrable Hamiltonian partial differential equations. The heights/speeds  of the solitions are governed by the eigenvalues of the associated isospectral eigenvalue problem. }
\label{peakon_figure}
\end{figure}  

In the absence of linear dispersion the CH equation (\ref{CH-eqn1}) is given by 
\begin{equation}
m_t+um_x+2mu_x=0
\quad\hbox{where}\quad
m=u-\alpha^2u_{xx}
\,,
\label{CH-eqn-intro}
\end{equation}
This limit of the CH equation retains only the effects of nonlinear dispersion on unidirectional shallow water waves and it is still completely integrable. The soliton solution for CH without linear dispersion is the {\bfi peakon} defined by
\begin{equation}\label{singlepeakon-soln}
u(x,t)=ce^{-|x-ct|/\alpha}
\,,
\end{equation}
which is the solitary traveling wave solution for (\ref{CH-eqn-intro}).
The peakon traveling wave moves at speed $c$ equal to its maximum height, at which it has a sharp peak (jump in derivative).  
The spatial velocity profile $e^{-|x|/\alpha}$ is the {\bfi Green's function} for the Helmholtz operator $(1 - \alpha^2\partial_{x}^2)$ on
the real line with vanishing boundary conditions at spatial infinity. In particular, it satisfies
\begin{equation}\label{Greens-relation}
(1 - \alpha^2\partial_{x}^2)e^{-|x-ct|/\alpha}
=
2\alpha \delta(x-ct)
\,,
\end{equation}
in which the {\bfi delta function} $\delta(x-q)$ is defined by
\index{delta function}
\begin{equation}
f(q) = \int f(x)\delta(x-q)\,dx
\,,
\label{delta-def}
\end{equation}
for an arbitrary smooth function $f$. 

In its dispersionless limit in (\ref{CH-eqn-intro}), the CH equation admits solutions representing a {\bfi wave train of peakons}
\begin{equation}\label{peakontrain-soln}
u(x,t)=\sum_{a=1}^Np_a(t)e^{-|x-q_a(t)|/\alpha}
\,.
\end{equation}
Such a sum is an \emph{exact solution} of the dispersionless CH equation (\ref{CH-eqn-intro}) provided the time-dependent parameters $\{p_a\}$ and $\{q_a\}$, $a=1,\dots,N$, satisfy certain canonical Hamiltonian equations that will be discussed later. 
By equation (\ref{Greens-relation}), the peakon wave train (\ref{peakontrain-soln}) corresponds to a sum over delta functions representing the {\bfi singular solution} in momentum,
\begin{equation}\label{m-delta-N}
m(x,t) = 2\alpha\sum_{a=1}^N p_a(t)\,\delta(x-q_a(t))
\,.
\end{equation}
A remarkable feature of these singular solutions is that they emerge from any spatially confined initial condition, as shown in the sequential plots of velocity profiles in Figure \ref{peakon_figure}. Being solitons with no internal degrees of freedom, the singular solutions interact by scattering elastically with each other.  These elastic-collision solution properties hold for any Green's function or convolution kernel $K(x)$ in the convolution relation $u=K*m$ between velocity $u$ and momentum $m$ \cite{FrHo2001}. The generalization 
\begin{equation}
m_t+um_x+2mu_x=0
\quad\hbox{where}\quad
u=K*m
\,,
\label{EPDiff-intro}
\end{equation}
for an arbitrary Green's function $K(x)$ is called {\bfi EPDiff}, because it arises as an Euler-Poincar\'e (EP) equation from Hamilton's principle for a Lagrangian defined as a metric on the tangent space of the diffeomorphisms (Diff) \cite{HoMaRa1998}. In particular, EPDiff describes geodesic motion on the diffeomorphisms with respect to the metric associated with the Green's function $K$.

The EPDiff formulation generalizes to higher spatial dimensions immediately.  In higher dimensions EPDiff arises in applications such as turbulence where it is the basis for the Navier-Stokes-alpha model \cite{FoHoTi01} and in imaging \cite{HoRaTrYo2004,HoTrYo2007} where it appears in the optimal control approach to template matching.
In any number of spatial dimensions, EPDiff admits the \emph{spontaneous} emergence of singular solutions from confined initial configurations.
In 1D, this emergent singular behavior results from the {\bfi steepening lemma}, proved for dispersionless CH with $K(x)=e^{-|x|/\alpha}$ in \cite{CaHo1993}. 

\subsection{Plan of the paper}
The dispersionless Camassa-Holm equation has a two-component integrable extension (CH2) \cite{ChLiZh2005,Ku2007,Falqui06}. The CH2 system of equations involves both fluid density and momentum, but it possesses singular solutions only in the latter variable. Section \ref{2component-sec} discusses CH2 in the context of shallow water systems. After discussing some of the shallow-water properties of CH2 in Section \ref{CH2-sec}, a modified system MCH2 is proposed in Section \ref{MCH2-sec}. Although the MCH2 system may not be integrable, it does allow delta-like singular solutions in {\it both} variables, not just the fluid momentum. These singular solutions exist in any number of dimensions, as discussed in Section \ref{singmommap-sec}. The mechanism of nonlinear steepening by which these singular solutions form is explained via a lemma in Section \ref{steep-sec}. 
This lemma reveals the conditions under which wave breaking occurs, so that the MCH2 fluid velocity in one spatial dimension develops a negative vertical slope, starting from smooth initial conditions. 
Section \ref{numerics-sec} presents numerical results of emergence and stability of MCH2 solutions for a typical fluid problem known as dam breaking. The numerical results indicate that the singular solutions are quite stable as they emerge and when they collide, but may tend toward the pure CH peakon solutions asymptotically in time. Section \ref{collidingpairs} analyzes the pairwise collisions of MCH2 peakons in one dimension. Section \ref{attractive-interact} addresses the numerical phenomenology of attractive interactions of peakons, in which the sign of the gravitational acceleration is opposite that of the shallow-water case. Section \ref{metamorph} concludes with some remarks about the potential applications of MCH2 in imaging science.

\section{Two-component systems}
\label{2component-sec}

\subsection{The two-component Camassa-Holm system (CH2)}
\label{CH2-sec}
In recent years, the Camassa-Holm (CH) equation (\ref{CH-eqn-intro}) has been extended so as to combine its integrability property with compressibility, or free-surface elevation dynamics in its shallow-water interpretation. This extension involves adding a continuity equation for the scalar density (or total depth) $\rho\in\mathcal{F}$ for real functions $\mathcal{F}$ and including a pressure term involving $\rho$ in the equation for the fluid momentum, as well as the fluid velocity $u$. 
The CH2 equations in that case are specifically  \cite{ChLiZh2005,Falqui06,Ku2007,ShAl84}
\begin{eqnarray}
m_t+um_x+2mu_x
&=&
-g\,\rho\rho_x
\quad\hbox{where}\quad
m=u-\alpha^2u_{xx}
\,,\nonumber\\
\rho_t +(\rho u)_x &=& 0
\,,
\label{CH2-eqns}
\end{eqnarray}
where $g>0$ is the downward constant acceleration of gravity in applications to shallow water waves. Boundary conditions are taken as $u\to0$ and  $\rho\to\rho_0=const$ as $|x|\to\infty$. 

Complete integrability of the CH2 system (\ref{CH2-eqns}) may be proven by writing it as the compatibility condition for two linear systems with a spectral parameter $\lambda$ as in \cite{ChLiZh2005,Ku2007,Falqui06}
\begin{eqnarray}
\psi_{xx}
&=& 
\Big( \frac{1}{4} 
+ m\lambda - g\rho^2\lambda^2\Big )\psi 
\quad\hbox{(Eigenvalue problem)}
\label{LaxPair-eigen}
\\
\psi_t
 &=& 
\Big( \frac{1}{2\lambda} - u\Big)\psi_x
+ \frac{1}{2}\,u_x\,\psi
\quad\hbox{(Evolution equation)}
\label{LaxPair-evol}
\end{eqnarray}
Requiring compatibility $\psi_{xxt}=\psi_{txx}$ and isospectrality $d\lambda/dt=0$ recovers the CH2 system (\ref{CH2-eqns}).

Geometrically, the CH2 system corresponds to geodesic motion with respect to the conserved metric \cite{HoTr2007, HoTrYo2007}
\begin{equation}
L(u,\rho)
=
\frac12\|u\|_{H^1}^2 + \frac{g}{2} \|\rho-\rho_0\|_{L^2}^2
\label{CH2-metric}
\end{equation}
for velocity and density $(u,\,\rho)\in T{\rm Diff}\,\times\,{\cal F}$, where $\mathcal{F}$ denotes the space of scalar functions. The cases $g>0$ (resp. $g<0$) correspond to repulsive (resp. attractive) interactions among particles of positive pointwise density $\rho$. The CH2 system emerges from Hamilton's principle $\delta S=0$ with the action $S=\int L(u,\rho)dt$ when the Lagrangian $L(u,\rho)$ is taken to be the metric in (\ref{CH2-metric}). 
The CH2 system (\ref{CH2-eqns}) has been shown to possess peakon solutions in velocity $u$ and corner-like solutions in density $\rho$ in \cite{ChLiZh2005}. However, singular solutions do not exist for its density variable \cite{CoIv2008}. Singular solutions in density $\rho$ (and therefore reduction to a finite-dimensional system of Hamiltonian equations) will be restored in the next section by a slight modification of CH2. 

\subsection{Modified CH2 (MCH2)}
\label{MCH2-sec}
The proposed modification of the CH2 system (\ref{CH2-eqns}) is expressed in terms of an averaged or filtered density $\rhobar$ in analogy to the relation between momentum and velocity by setting
\rem{
\begin{equation}
u=K_1*m
\quad\hbox{and}\quad
\rhobar =K_2*\rho
\,,
\label{avg-variables}
\end{equation}
where $K_1$ and $K_2$ are Green's functions for Helmholtz operators with potentially different length scales $\alpha_1$ and $\alpha_2$, so that
}
\[
m=(1-\alpha^2_1\partial^2)\, u
\quad\hbox{and}\quad
\rho=(1-\alpha^2_2\partial^2)\, \rhobar
\,,
\]
where one denotes $\partial = \partial / \partial  x$ and defines two length scales $\alpha_1$ and $\alpha_2$.
This modification will amount to strengthening the norm for $\rhobar$ from $L^2$ to $H^1$ in the potential energy term in the metric Lagrangian (\ref{CH2-metric}) in Hamilton's principle for the CH2 system. 
\\[2mm]

The modified CH2 system (MCH2) is written in terms of velocity $u$ and locally-averaged density $\rhobar$ (or depth, in the shallow-water interpretation).  MCH2 is defined as geodesic motion on the semidirect product Lie group Diff$\,\circledS\,\mathcal{F}$ with respect to a certain metric and is given as a set of Euler-Poincar\'e equations on the dual of the corresponding Lie algebra $\mathfrak{X}^{\,}\circledS\,\mathcal{F}$. In the general case, for a Lagrangian $L(u,\rhobar)$, the corresponding semidirect-product Euler-Poincar\'e equations are written as \cite{HoMaRa1998}
\begin{eqnarray}
\frac{\partial}{\partial  t}\frac{\delta L}{\delta u}
=
-\,\pounds_{u\,}\frac{\delta L}{\delta u}
-\frac{\delta L}{\delta \rhobar}\,\nabla \rhobar
\,,\qquad
\frac{\partial}{\partial  t}\frac{\delta L}{\delta \rhobar}
=-\,\pounds_u\,\frac{\delta L}{\delta \rhobar}
\,,
\label{EP-eqns}
\end{eqnarray}
where $\pounds_u\,(\delta L/\delta u)$ is the Lie derivative of the one-form density $m=\delta L/\delta u$ with respect to the vector field $u$ and $\pounds_u\,\delta L/\delta \rhobar$ is the corresponding Lie derivative of the scalar density $\delta L/\delta \rhobar$.
\\[1mm]

The integrable CH2 system and thee modified system MCH2 may both be derived as semidirect-product Euler-Poincar\'e equations (\ref{EP-eqns}) from the following type of variational principle defined
on the Lie algebra $\mathfrak{X}^{\,}\circledS\,\mathcal{F}$
\[
\delta\!\int_{t_0}^{t_1}\!\!L(u, \rhobar)\,{\rm d}t=0
\,,
\]
with Lagrangian
\begin{eqnarray}
L(u, \rhobar)
&=&
\frac12\|u\|_{H^1}^2 
+
\frac{g}{2}\|\rhobar-\rhobar_0\|_{H^1}^2
\label{MCH2-norm}
\\&=&
\frac12\int \!(u^2 + \alpha_1^2u_x^2)\,{\rm d}x
+
\frac{g}{2}\int \Big[(\rhobar -\rhobar_0)^2
+
\alpha_2^2\,(\rhobar-\rhobar_0)_x^2
\Big]
\,{\rm d}x
\,,
\label{MCH2-Lag}
\end{eqnarray}
in which the last line defines the $H^1$ norms of $u$ and $(\rhobar-\rhobar_0)$ and $\rhobar_0$ is taken to be constant.
The variational derivatives of this Lagrangian define the variables $m$ and $\rho$ as
\begin{eqnarray*}
\frac{\delta L}{\delta \rhobar}  
&=& 
g(1 - \alpha_2^2\partial^2)(\rhobar - \rhobar_0)
=:
g\rho 
\,,\\
\frac{\delta L}{\delta u}  
&=& 
(1 - \alpha_1^2\partial^2)u
=:m
\,.
\end{eqnarray*}
Substituting these variational derivatives into the Euler-Poincar\'e equations (\ref{EP-eqns}) recovers the CH2 equations (\ref{CH2-eqns}) for the constant values $g>0$, $\alpha_1 = \alpha$ and  $\alpha_2 =0$. However, when $\alpha_2^2 >0$ the MCH2 equations result. Namely, in one dimension the MCH2 system is 
\begin{eqnarray}
m_t+um_x+2mu_x
&=&
-\,g\,\rho\rhobar_x
\,,\nonumber\\
\rho_t +(\rho u)_x &=& 0
\,,
\label{MCH2-eqns}
\end{eqnarray}
where $m$ and $\rho$ are defined in terms of $u$ and $\rhobar$ via the Helmholtz operators
\begin{equation}
m:=(1 - \alpha_1^2\partial^2)u
\quad\hbox{and}\quad
\rho:=(1 - \alpha_2^2\partial^2)(\rhobar - \rhobar_0)
\,.
\label{m-rho-defs}
\end{equation}
This slight modification of CH2 defines the MCH2 equations in one dimension and, as we shall show, it suffices for the existence of emergent singular solutions in both $m$ and $\rho$. 

\begin{remark}[CH2 and MCH2 in higher dimensions]
In higher dimensions the Lagrangian (\ref{MCH2-Lag}) generalizes to
\begin{equation}
L({\bf u}, \rhobar)
=
\frac12\int \!{\bf u}\cdot \ (1-\alpha_1^2\Delta) {\bf u}\,\,{\rm d}^n{\bf x}
+\frac{g}{2}\int (\rhobar-\rhobar_0)
(1-\alpha_2^2\Delta)\, (\rhobar-\rhobar_0)\,{\rm d}^n{\bf x}
\,,
\label{nDCH-Lag}
\end{equation}
where $\Delta$ is the $n$-dimensional Laplacian operator. 
The semidirect-product Euler-Poincar\'e equations for this Lagrangian produce the desired generalizations of CH2 and MCH2 to higher dimensions.
\end{remark}

\rem{
Upon specializing the operator $Q_2$, this system may have the physical interpretation of an {\it incompressible charged fluid}. In fact, we can think of $\rho$ as a charge (rather than mass) density. Then, in the special case $Q_2=\Delta$ we have the pure electrostatic interaction $\Delta\Phi=\rho$. The pure CH-2 case ($Q_2=1$) corresponds to a delta-like interaction potential, which also appears in the integrable Benney system \cite{Be1973,Gi1981}.
}

\section{Existence of singular solutions}
\label{singmommap-sec}
One evaluates the variational derivatives of the $n$-dimensional Lagrangian (\ref{nDCH-Lag}) as
\begin{equation}
\frac{\delta L}{\delta u}
=
\mathbf{m}\cdot {\rm d}\mathbf{x}\otimes{\rm d}V
\quad\hbox{and}\quad
\frac{\delta L}{\delta \rhobar} = g\rho\otimes {\rm d}V
\,,
\label{var-coords}
\end{equation}
with co-vector $\mathbf{m}$ and scalar function $\rho$ defined by 
\begin{equation}
\mathbf{m}
:=
(1-\alpha_1^2\Delta)\mathbf{u}
\quad\hbox{and}\quad
\rho = (1-\alpha_2^2\Delta)\rhobar
\,.
\label{var-coords}
\end{equation}
Then the semidirect-product Euler-Poincar\'e equations (\ref{EP-eqns})  in $\mathbb{R}^n$ may be written in coordinates as
\begin{eqnarray}
\frac{\partial \mathbf{m}}{\partial t}
&=&
- 
\underbrace{\
\mathbf{u}\cdot\nabla \mathbf{m}\
}_{\hbox{Convection}}
-\ 
\underbrace{\
(\nabla \mathbf{u})^T\cdot\mathbf{m}\
}_{\hbox{Stretching}}\
-\ 
\underbrace{\
\mathbf{m}({\rm div\,}\mathbf{u})\
}_{\hbox{Expansion}}
-\
\underbrace{\
g\rho\nabla\rhobar\
}_{\hbox{Force}}
\,,
\label{sdEP-mom-eqn}\\
\frac{\partial \rho}{\partial t} 
&=&
-\   {\rm div}\,\rho\mathbf{u} 
\,.
\label{sdEP-dens-eqn}
\end{eqnarray}
Here one denotes $(\nabla \mathbf{u})^T\cdot\mathbf{m}=\sum_j m_j\nabla u^j$.
To explain the terms in underbraces, we rewrite EPDiff as the change in the one-form density of momentum along the characteristic curves of the velocity. In vector coordinates, this is 
\begin{equation}\label{EPDiff-char-form}
\frac{d}{dt}\Big(\mathbf{m}\cdot d\mathbf{x}\otimes {\rm d}V\Big)
=
-\,g\rho d\rhobar\otimes {\rm d}V
\quad\hbox{along}\quad
\frac{d\mathbf{x}}{dt}=\mathbf{u}=K_1*\mathbf{m}
\,.
\end{equation} 
This form of the EPDiff equation also emphasizes its nonlocality, since the velocity is obtained from the momentum density by convolution against the Green's function $K_1$ of the first Helmholtz operator in (\ref{var-coords}). Likewise, the average density $\rhobar$ follows from the pointwise density $\rho$ by convolution against the Green's function $K_2$.

A direct substitution into the Euler-Poincar\'e equations (\ref{sdEP-mom-eqn}) and (\ref{sdEP-dens-eqn}) in $\mathbb{R}^n$ coordinates shows that these equations possess the following singular solutions
\begin{equation}
\big({\bf m},\rho\big)=\sum_{i=1}^N\int\!\big({\bf P}_i(s,t),w_i(s)\big)\,\delta\!\left({\bf
x-Q}_i(s,t)\right)\,{\rm d}^ks
\,,
\label{nDsingsoln}
\end{equation}
where (${\bf Q}_i,{\bf P}_i$) satisfy
\begin{align*}
\pp{{\bf Q}_i}{t}=&\sum_j\int {\bf P}_j(s',t)\ K_1\!\left({\bf Q}_i(s,t)-{\bf Q}_j(s',t)\right)\,{\rm d}^ks'
\,,\\
\pp{{\bf P}_i}{t}=&-
\,\sum_j\int \left({\bf P}_i(s,t)\cdot{\bf P}_j(s',t)\right) \pp{}{{\bf
Q}_i}K_1\!\left({\bf Q}_i(s,t)-{\bf Q}_j(s',t)\right)\,{\rm d}^ks'
\\
&-
\sum_j\int w_i(s)\,w_j(s')\, \pp{}{{\bf
Q}_i}K_2\!\left({\bf Q}_i(s,t)-{\bf Q}_j(s',t)\right)\,{\rm d}^ks'
\,,
\end{align*}
where $K_1$ and $K_2$ are the kernels for the Green's functions of the Helmholtz operators with length scales $\alpha_1$ and $\alpha_2$, respectively. The continuity equation (\ref{sdEP-dens-eqn}) implies that the weights $w_i$ are independent of time. These equations may be written as Hamilton's canonical equations with Hamiltonian
\begin{multline*}
H_N\, =\, \frac12\sum_{i,j}^N\iint{\bf P}_i(s,t)\cdot{\bf P}_j(s',t)\ K_1\!\left({\bf Q}_{i}(s,t)-{\bf Q}_j(s',t)\right)\,{\rm d}^ks\,{\rm d}^ks'
\\
+
\frac12\sum_{i,j}^N\iint w_i(s)\,w_j(s')\, K_2\!\left({\bf Q}_i(s,t)-{\bf Q}_j(s',t)\right)\,{\rm d}^ks\,{\rm d}^ks'
\,.
\end{multline*}

\begin{remark}[Why are these equations canonical?]
The explanation of why the dynamical equations for the singular solutions are canonically Hamiltonian is beyond the scope of the present paper. This is explained from the perspective of Lie group actions and momentum maps in \cite{HoTr2007}.
\end{remark}

In the one dimensional treatment considered in the remainder of this work, the equations above simplify to
\begin{align}\nonumber
\pp{Q_i}{t}=&\sum_j\, {P}_j(t)\ K_1\!\left({Q}_i(t)-{Q}_j(t)\right)
\,,\\
\pp{P_i}{t}=&-
\,\sum_j\,{P}_i(t)\,{P}_j(t)\ \pp{}{{Q}_i}
K_1\!\left({Q}_i(t)-{Q}_j(t)\right)
-
\sum_j\, w_i\,w_j\ \pp{}{{Q}_i}
K_2\!\left({Q}_i(t)-{Q}_j(t)\right)
\label{collective}
\,,
\end{align}
and if we choose $\alpha_2=1$ so that $K_2(x)=\frac12e^{\,-|x|}$, we recover the peakon solution
for $\rhobar $
\[
\rhobar(x,t)=(1-\partial_x^2)^{-1}\rho
=\frac12\sum_i\,w_i\,e^{\,-\left|x-Q_i(t)\right|}
\,,
\]
as well as the usual peakon wave train solution (\ref{peakontrain-soln}) for the velocity $u$.

\begin{remark}[Singular potential terms and integrable cases]
In the limiting case \makebox{$\alpha_2=0$}, one recovers the CH2 system corresponding to geodesic motion with a delta-function interaction potential. In terms of particle solutions, this means that for a positive potential $(g>0)$, two particles will bounce off immediately before colliding; while for a negative potential $(g<0)$, they will proceed together, attached one to the other and they will never split apart. This kind of singular delta-like potential is also present in another integrable system, namely the Benney equation \cite{Be1973,Gi1981}.
\end{remark}

\section{Steepening lemma for MCH2}
\label{steep-sec}
Do the singular MCH2 solutions emerge from smooth initial conditions? This is not automatic. For instance, we know that point vortices are invariant singular solutions that are not generated by vorticity dynamics with smooth initial conditions, whereas peakon solutions are always {\it spontaneously} generated by the CH flow. 

This section and the next address the following two questions:
\begin{enumerate}
\item Under what conditions does MCH2 admit emergent singular solutions?
\item What are the stability properties of such solutions?
\end{enumerate}
This section proves a {\bfi steepening lemma} which reveals the conditions under which the singular solutions for MCH2 in one spatial dimension emerge from smooth initial conditions.

On using the definitions of $m$ and $\rho$ in (\ref{m-rho-defs}), the MCH2 equations (\ref{MCH2-eqns}) may be rewritten in a form that transparently displays its nonlocal nature 
\begin{eqnarray}
u_t+uu_x
&=&
-\,\partial\, (K_1*p)
\quad\hbox{with}\quad
p:=
u^2 
+ \frac{\alpha_1^2}{2}\,u_x^2
+ \frac{g}{2}\,\rhobar^2
- \frac{g\alpha_2^2}{2}\,\rhobar_x^2
\,,
\label{MCH2-u}
\\
\rhobar_t + u \rhobar_x 
&=& 
-\,K_2*
\Big( \alpha_2^2\, \big(u_x\rhobar_x \big)_x
+
(u_x\rhobar)
\Big)
\,.
\label{MCH2-rhobar}
\end{eqnarray}
where $K_1$ and $K_2$ are the Green's functions for the corresponding Helmholtz operators. For $g>0$, conservation of the $H^1$ norm (\ref{MCH2-norm}) ensures that $u$ and $\rhobar$ are bounded pointwise, because of the Sobolev inequality valid in one dimension that, 
\begin{equation}
\max_{x\in\mathbb{R}}\Big(u^2(x,t)
+
g \big(\rhobar(x,t)-\rhobar_0 \big)^2 \Big)
\le
\rem{
\int_{-\infty}^\infty
\left(u^2+\alpha_1^2u_x^2\right)
+
g(\rhobar-\rhobar_0)^2+\alpha_2^2 (\rhobar-\rhobar_0)_x^2)dx
}
\|u\|_{H^1}^2+g\|\rhobar-\rhobar_0\|_{H^1}^2
=const
\,.
\label{Sobolev-bdd}
\end{equation}

\begin{lemma}[Steepening Lemma for the MCH2 system]$\quad$\\
\label{steep-lemma}
Suppose the initial profile of velocity $u(0,x)$ has an inflection point at
$x=\overline{x}$ to the right of its maximum, and otherwise it decays to zero in each direction. Under MCH2 dynamics with $g>0$ a sufficiently negative slope at the inflection point will become vertical in finite time.
\end{lemma}

\begin{proof}
Consider the evolution of the slope
\[
s(t) =u_x(\overline{x}(t),t)
\,,
\]
at the inflection point $x=\overline{x}(t)$.
The spatial derivative of the $u$-equation  (\ref{MCH2-u}) yields an equation for the evolution of $s$. Namely, using $u_{xx}(\overline{x}(t),t)=0$ the spatial derivative leads to 
\begin{eqnarray} \label{b-slope-eqn1}
\frac{ds}{dt} + s^2 
&=&
-\,\partial_x^2
(K_1*p)
\quad\hbox{with}\quad
p:=\bigg[u^2 
+ \frac{\alpha_1^2}{2}\,s^2
+ \frac{g}{2}\,\rhobar^2
- \frac{g\alpha_2^2}{2}\,\rhobar_x^2 \bigg]_{x=\overline{x}(t)}
\nonumber\\
&=&
\frac{1}{\alpha_1^2}(1-\,\alpha_1^2\partial_x^2)K_1*p
-\frac{1}{\alpha_1^2}K_1*p
\nonumber\\
&=&
\frac{1}{\alpha_1^2}p - \frac{1}{\alpha_1^2}K_1*p 
\,.
\end{eqnarray}
This calculation implies with $g>0$
\begin{eqnarray} \label{slope-eqn2}
\frac{ds}{dt} 
&=& 
-\,\frac{1}{2} s^2
-\frac{1}{\alpha_1^2}\int_{-\infty}^\infty
\,e^{-|\overline{x}-y|/\alpha_1}
\left(u^2 + \frac{g}{2}\,\rhobar^2
- \frac{g\alpha_2^2}{2}\,\rhobar_y^2\right)dy
+ 
\frac{1}{\alpha_1^2}
\bigg[u^2 
+ \frac{g}{2}\,\rhobar^2
- \frac{g\alpha_2^2}{2}\,\rhobar_x^2 \bigg]_{x=\overline{x}(t)}
\nonumber\\
&\le&  
-\,\frac{1}{2} s^2 
+
\frac{g\alpha_2^2}{2\alpha_1^2}\int_{-\infty}^\infty
\,e^{-|\overline{x}-y|/\alpha_1}
\,\rhobar_y^2\,dy
+
\frac{1}{2\alpha_1^2}
\bigg[u^2 
+ \frac{g}{2}\,\rhobar^2 \bigg]_{x=\overline{x}(t)}
\nonumber\\
&\le&  
-\,\frac{1}{2} s^2 
+
\frac{g\alpha_2^2}{\alpha_1^2}\int_{-\infty}^\infty\!
\,(\rhobar-\rhobar_0)_x^2\,dx
+
\frac{1}{\alpha_1^2}
\bigg[u^2 
+ \frac{g}{2}\,\rhobar^2 \bigg]_{x=\overline{x}(t)}
\,,
\label{slope-estimate}
\end{eqnarray}
where the second step drops the negative terms in the previous line and the last step uses $e^{-|x-y|}\le1$. By conservation of the $H^1$ norm with $g>0$ in the Sobolev bound (\ref{Sobolev-bdd}), the sum of last three terms remains finite, say less than a number $M/2$. Consequently, we have
\begin{eqnarray} \label{slope-eqn3b}
\frac{ds}{dt} 
&\le& -\,\frac{1}{2} s^2 +  \frac{M}{2}
\,,
\end{eqnarray}
which implies, for a sufficiently negative initial slope, $s\le-\sqrt{M}$, that the slope remains negative and becomes vertical in finite time, as 
\begin{eqnarray} \label{slope-eqn4}
s(t) 
&\le&\sqrt{M}\coth\left(\sigma + \frac{t}{2}\sqrt{M}\right)
\,,
\end{eqnarray}
where $\sigma$ is a negative constant that determines the initial slope, also negative.
Hence, at time $t=-2\sigma/\sqrt{M}$ the slope becomes negative and vertical. This wave-breaking result for the slope of the fluid velocity proves the steepening lemma for the MCH2 equation with $g>0$. 
\end{proof}

\begin{remark}
The calculation corresponding to (\ref{slope-estimate}) for $g<0$ results in the estimate
\[
\frac{ds}{dt}
\le
-\,\frac{1}{2} s^2 
+
\frac{|g|\alpha_2^2}{\alpha_1^2}\int_{-\infty}^\infty\!
\,\rhobar^2\,dx
+
\frac{1}{\alpha_1^2}
\bigg[u^2 
+ 
\frac{|g|\alpha_2^2}{2}\,\rhobar_x^2\bigg]_{x=\overline{x}(t)}
\,.
\] 
However, without any control on the pointwise value of $\rhobar_x^2$, no real conclusion may be drawn from this estimate. 
\end{remark}

\begin{remark}[Density at the point of wave breaking]
One may ask how the density behaves when the velocity slope at the inflection point becomes vertical in finite time for $g>0$. This may be determined by integrating the continuity equation for $\rho$ along a Lagrangian path $x=\chi(x_0,t)$ with $\chi(x_0,0)=x_0$. Evaluating at the inflection point yields
\[
\bigg[
\frac{1}{\rho(x_0,t)}
\frac{d\rho(x_0,t)}{dt}
\bigg]_{x(x_0,t)=\overline{x}(t)} 
= -\,s(t)
\,,
\]
so that the density of the Lagrangian parcel currently occupying position $\overline{x}(t)$ is given by
\[
\rho(x_0,t)\Big|_{x(x_0,t)=\overline{x}(t)}
=
\rho(x_0,0)\exp \bigg(-\int_0^t s(\tau)d\tau \bigg)
\,.
\]
Because the blow up in $s(t)$ as $\coth$ in equation (\ref{slope-eqn4}) is integrable, the density of the Lagrangian parcel at the inflection point remains finite even though the velocity develops a vertical slope. Thus, wave breaking in the fluid velocity does not imply singularity in the pointwise density at the point of vertical slope.  
\end{remark}

\begin{remark}[The formation of delta singularities in momentum]
$\quad$\\
Peakons have no inflection points, so they are exempt from the steepening lemma.  As shown in Figure \ref{peakon_figure}, their formation creates a new inflection point of less negative slope. The negative slope of the inflection point present in the initial velocity profile steepens and the velocity at the inflection point rises until it lifts above the initial maximum velocity. Then it propagates away, leaving an inflection point behind, which will soon repeat the process. At each peak in the velocity profile the slope takes a jump, so the curvature there is infinite. This in turn leads to the delta singularities in the momentum variable $(m)$ obtained by applying the Helmholtz operator to the velocity profile. 
Finally, the formation of the delta-function singularity in momentum  $m$ must also be accompanied by a delta-function singularity in pointwise density $(\rho)$, as shown by direct substitution into the MCH2 equations in Section \ref{singmommap-sec}. Thus, the reaction to wave breaking by nonlinear steepening at inflection points of negative slope creates peakons and is the mechanism under which MCH2 forms its delta-singularities in $m$ and $\rho$. 
\end{remark}

\section{Numerical results benchmark problem: dam breaking and bores}
\label{numerics-sec}
The simulations reported in this section solve the following equations with $g=1$
\rem{
\begin{eqnarray}
\partial_t \rhobar
&=&
\left(1-\alpha_2^2\partial_{xx}\right)^{-1}\partial_x\left(-u \rhobar +\alpha_2^2u\partial_{xx} \rhobar\right),\nonumber\\
\partial_t u&=&\left(1-\alpha_1^2\partial_{xx}\right)^{-1}\left(-3u\partial_x
u+2\alpha_1^2\partial_x u\partial_{xx}u+\alpha_1^2u\partial_{xxx}u-\rhobar\partial_x \rhobar +\alpha_2^2\partial_x \rhobar\partial_{xx} \rhobar\right),
\end{eqnarray}
}
\begin{eqnarray}
\partial_t u
&=&
K_1*\left(-3u u_x
+2\alpha_1^2 u_x u_{xx}
+\alpha_1^2u u_{xxx}
-\,g(\rhobar  - \alpha_2^2 \rhobar_{xx})\rhobar_x\right)
\,,\nonumber\\
\partial_t \rhobar
&=&
-
K_2*\partial\left( (\rhobar -\alpha_2^2 \rhobar_{xx})u \right)
\,,
\label{rhobaru-eqns}
\end{eqnarray}
where $\alpha_1$ and $\alpha_2$ are the two length scales of nonlocality. The first equation may also be written in conservative form as
\[
u_t+u u_x = -K_1*\partial\left(u^2+\frac{\alpha_1^2}{2}u_x^2
+\frac{g}{2} \rhobar^2-\frac{g\alpha_2^2}{2}\rhobar_x^2\right)
\,.
\]
Periodic boundary conditions are imposed on the domain $\left[-L,L\right]$ with {\bfi dam-break initial conditions} given by
\begin{equation}
u\left(x,0\right)=0,\qquad \rhobar\left(x,0\right)
= 1 +  \tanh(x+a)-\tanh(x-a) 
\,,
\label{dambreak-ic}
\end{equation}
where $a\ll L$.

\subsection{Dam breaking for $g>0$}

The dam-break arises when a body of water of uniform depth is
retained behind a barrier, in this case at $x=\pm a$.  If this barrier is
suddenly removed at $t=0$, then the water flows under gravity.  The problem
is to find the subsequent flow and moreover, to determine the shape of the
free surface.  This question is addressed in the context of shallow-water
theory by Acheson (see~\cite{Ach1990}), and thus serves as a typical hydrodynamic
problem to be discussed in our framework.

In the following paragraphs, we compare the numerical results obtained in two different cases: the exactly integrable CH2 system; and the modified MCH2 equations (\ref{MCH2-eqns}) proposed above (with $\alpha_2\neq0$). 

\begin{figure}[htb]
\subfigure[]{
\includegraphics[width=0.48\textwidth]{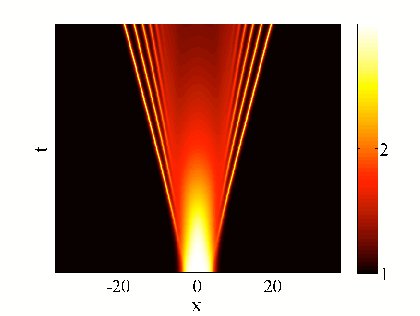}
}
\subfigure[]{
\includegraphics[width=0.48\textwidth]{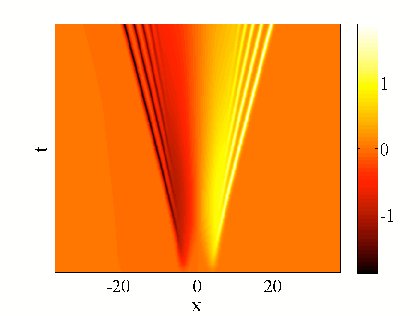}
}
\caption{(Color online) Dam-break results for the pure CH2 system show (a)  evolution of the density $\rho= \rhobar $ and (b) evolution of the velocity $u$ from initial conditions (\ref{dambreak-ic}).  } 
\label{fig:pde_alpha0}
\end{figure}

\paragraph{Case 1: pure CH2 equations.} 
We take $\alpha_1=0.3$, $\alpha_2=0$, $a=4$, and $L=12\pi$.
The choice $\alpha_2=0$ gives a density equation (the equation in $\rhobar $)
that is unsmoothed. As shown in Figure \ref{fig:pde_alpha0}, the singular solutions emerge after finite time, and both variables have this singular property.
More and more singular elements appear as time progresses.  

\rem{
However, the singular density elements decay to zero, while the singular velocity elements persist. The plot on the left side shows that $\rhobar $-singularities that are created later tend to  persist longer.
}

\begin{figure}[h!]
\subfigure[]{
\includegraphics[width=0.48\textwidth]{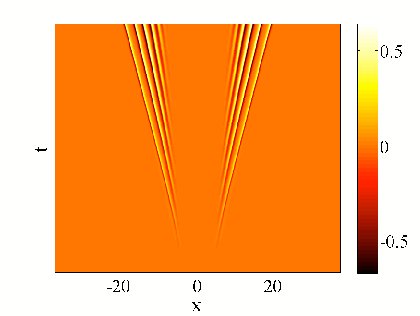}
}
\subfigure[]{
\includegraphics[width=0.48\textwidth]{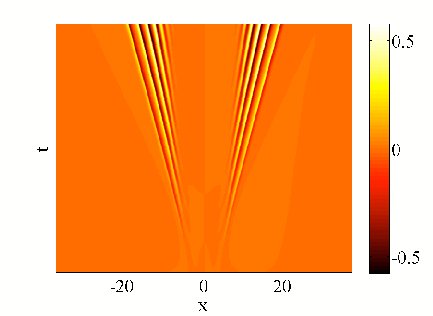}
}
\caption{(Color online) The \emph{appearances} of the dam-break results for the MCH2 and CH2 systems are identical. See Figure \ref{fig:pde_alpha0}. The \emph{differences} plotted in this figure show (a)  evolution of the filtered density difference $\rhobar - \rhobar_0$ and (b) evolution of the velocity difference $u-u_0$. Here the subscript zero refers to CH2, for which $\alpha_2=0$.}
\label{fig:pde_diff_alpha}
\end{figure}

In contrast, one may compare the evolution of the velocity $u$ for the single-component CH equation launched from a spatially confined \emph{velocity profile}: $u(x,0)=\tanh(x+4)-\tanh(x-4)$ and $\rhobar(x,0)={const}$ with $\alpha_1=0.3$. In contrast to CH2, only rightward-moving peakons emerge from CH as illustrated in Figure \ref{fig:pde_alpha0}. Thus, the limit $\alpha_2\to0$ breaks the left-right symmetry of the CH2 system. 
\begin{figure}[htb]
\begin{center}
\includegraphics[width=0.48\textwidth]{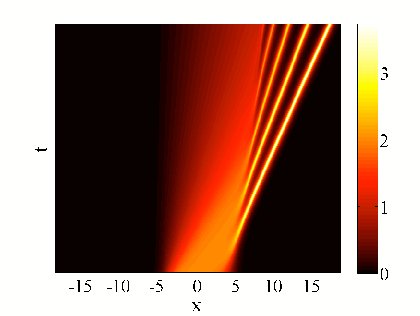}
\caption{(Color online) Evolution of the velocity $u$ for single-component CH from the spatially confined initial velocity profile, $u\left(x,0\right)=\tanh(x+a)-\tanh(x-a)$. } 
\label{fig:pde_alphaCH1}
\end{center}
\end{figure} 
\paragraph{Case 2: MCH2.}  We take $\alpha_1=0.3$, $\alpha_2=0.1$, $a=4$, and $L=12\pi$. The solution behavior in this case for the modified system with small $\alpha_2$ shows only slight difference from the pure CH2 case, as seen in Figure \ref{fig:pde_diff_alpha}. One difference is that now the singularities have a particle-like nature, whose collective dynamics can be studied using Hamilton's canonical equations (\ref{collective}).

\begin{figure}[h!]
\begin{center}
\subfigure[]{
\includegraphics[width=0.45\textwidth]{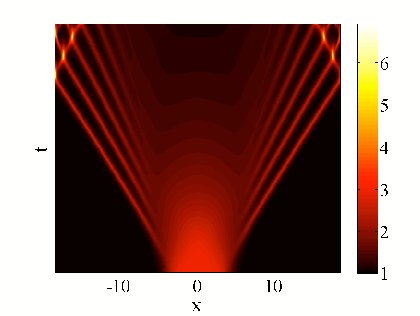}
}
\subfigure[]{
\includegraphics[width=0.45\textwidth]{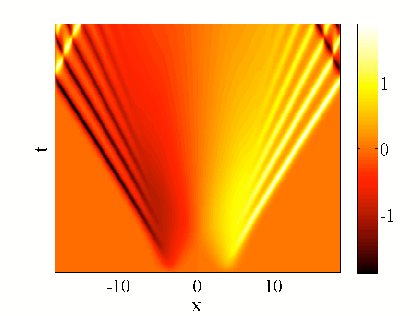}
}
\end{center}
\caption{(Color online) Reducing the size of the periodic domain for the dam-break for the MCH2 system allows the peakon wave train to wrap around, thereby allowing the leading peakons to suffer head-on collisions with the trailing ones. One observes a phase shift, during which the peakons form an unstable intermediate state, which then splits into peakons again. This phase shift is discussed further in Section \ref{collidingpairs}.
}
\label{fig:pde_headonMCH2}
\end{figure}

\begin{remark}
One might question whether the Steepening Lemma \ref{steep-lemma} for MCH2 applies to the dam-breaking initial conditions, because a zero-velocity initial condition has no inflection points. However, the numerics shows that as the dam breaks the right-hand-side of MCH2 immediately generates nonzero velocity. The velocity must develop at least one inflection point of negative slope to remain confined with finite speed. This situation qualifies it for the lemma. The lemma says that once an inflection point of negative slope develops, wave breaking has a deadline. The MCH2 equation avoids wave breaking by successively forming peakons, each time shifting the inflection point backward before its deadline and restarting its clock at the new location.
\end{remark}

\rem{
The fact that the $\rhobar $-singularities decay to zero might be unexpected. Indeed, since these correspond to particle dynamics, such solutions form an invariant manifold and they would be expected to persist after their formation. On the other hand, the MCH2 dynamics  is governed by a system of equations, which can evolve to some asymptotic state. Thus, the simulations show that the nonzero weight solutions are unstable to going to zero, whereupon they become stable.

Another objection could concern conservation of mass/charge, since the average density carried by some fluid particles seem to evanesce after their creation. However, this evanescence of mass carried by one particle  seems to be compensated by the simultaneous creation of another $\rhobar $-peakon, which balances the loss. The numerical results above have been checked to ensure that they accurately respect overall mass conservation.
}

\subsection{Bore}

An alternative to the stationary dam breaking problem is to consider the dam breaking initial condition as moving in a frame of uniform motion and regard the result as modelling the development of a bore  produced by a moving elevation of water. 

Equations (\ref{rhobaru-eqns}) are solved with $g=1$ and the following alpha-values: $\alpha_1=0.3$ and $\alpha_2=1$, together with the initial conditions
\[
u\left(x,0\right)=\tanh\left(x+4\right)-\tanh\left(x-4\right),\qquad
\rhobar\left(x,0\right)=1.
\] 
The larger value of $\alpha_2$ was needed to prevent gradients in $\rhobar$ from becoming nearly vertical for the chosen resolution.
Gradients in $\rhobar$ are not a problem for the dam-break initial conditions because the trailing density peaks in that case remain small in magnitude. However, in the bore configuration the density builds up as time proceeds. Since a large value of $\alpha_2$ was needed to control the gradients in $\rhobar$ for the bore, a comparison with the $\alpha_2=0$ case was not possible.

\begin{remark}
Similar results are obtained for various other initial density profiles. For example, with $\rhobar\left(x,0\right)=b\left[\tanh\left(x+4\right)-\tanh\left(x-4\right)\right]$ and $b=0.1$, the density profile remains finite, and the velocity profile is very similar to that in Figure \ref{fig:pde_alphaCH1} for CH.  Similarly, for $\rhobar\left(x,0\right)=0$, the density profile remains zero for all time, and the velocity profile again closely resembles that in Figure \ref{fig:pde_alphaCH1}. 
\end{remark}

\begin{figure}[h!]
\subfigure[]{
\includegraphics[width=0.48\textwidth]{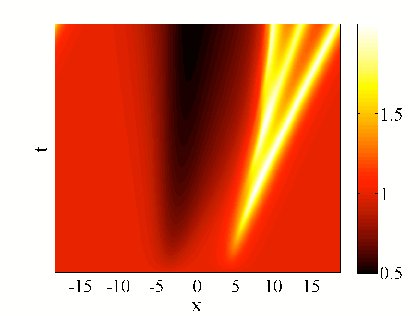}
}
\subfigure[]{
\includegraphics[width=0.48\textwidth]{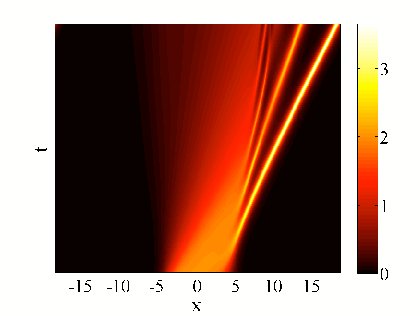}
}
\caption{(Color online) (a)  Space-time evolution of the function $\rhobar $; (b) Space-time evolution of the velocity $u$ for the MCH2 system in the bore configuration. Figure \ref{fig:pde_alpha}b for the velocity of the MCH2 system is quite similar to Figure \ref{fig:pde_alphaCH1} for CH, but close inspection shows that it is not identical.
}
\label{fig:pde_alpha}
\end{figure} 

\section{Pairwise collisions in one dimension for $g>0$}
\label{collidingpairs}
\begin{figure}[htb]
  \begin{center}
\includegraphics[width=0.5\textwidth]{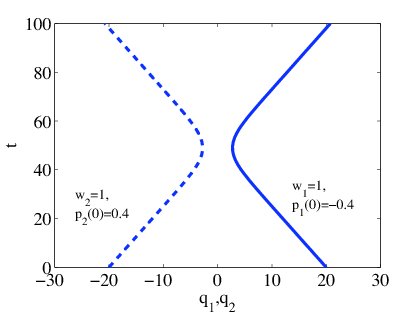}
  \end{center}
\caption{Particle trajectories for $\be_1\cdot\be_2>0$.}
\label{fig:overtaking}
\end{figure}

The interaction of two singular solutions may be
analyzed by truncating the sums in (\ref{nDsingsoln})  to consider $N=2$. In one dimension this yields
\[
H_2=
\frac12 \Big(\left.P_1^{2}+P_2^{2}
+
2\left.K_{1}(Q_1-Q_2\right)P_1\,P_2
\Big)\right.
+
\frac{g}{2} \Big(\left.w_1^{2}+w_2^{2}
+
2\left.K_{1}(Q_1-Q_2\right)\,w_1\,w_2
\Big)\right.
\]
and we choose $g=1$, corresponding to repulsive interactions. 
Following \cite{HoSt03}, one defines
\begin{align*}
P&=P_1+P_2
\,;\quad
Q=Q_1+Q_2
\,;\quad
p=P_1-P_2
\,,\quad
q=Q_1-Q_2
\,;\quad
W=w_1+w_2
\,;\quad
w=w_1-w_2
\,.
\end{align*}
Consequently, the Hamiltonian can be written as
\[
\mathcal{H}=\frac12(P^2+W^2)-\frac14\Big[(P^2-p^2)\left(1-K_1(q)\right)
+(W^2-w^2)\left(1-K_2(q)\right)\!\Big]
\,.
\]
At this point one writes the canonical equations
\begin{align}
\frac{dP}{dt}&=-2\frac{\partial \mathcal{H}}{\partial Q}=0
\,,
\,
\hspace{7.7cm}
\frac{dQ}{dt}=2
\frac{\partial \mathcal{H}}{\partial P}=P\left(1+K_1(q)\right)
\,,\nonumber\\
\frac{dp}{dt}&=
-2\frac{\partial \mathcal{H}}{\partial q}=
-\,\frac12\Big[\!\left(P^2-p^2\right)K_1^{\,\prime}(q)
+\left(W^2-w^2\right)K_2^{\,\prime}(q)\Big]
\,,
\hspace{1cm}
\frac{dq}{dt}=
2\frac{\partial \mathcal{H}}{\partial p}=
p\left(1-K_1(q)\right)
\,,
\label{eq:ode}
\end{align}
with $\dot{W}=\dot{w}=0$. These equations possess the first integral,
\[
\left(\frac{dq}{dt}\right)^2=
P^2\left(1-K_1(q)\right)^2
-
\big[\,4\mathcal{H}-2W^2+\left(W^2-w^2\right)\left(1-K_2(q)\right)\big]
\left(1-K_1(q)\right)
\,,
\]
which finally produces the quadrature
\[
dt=\frac{dK_1}{K_1^{\,\prime}\sqrt{P^2\left(K_1^{\,2}(q)-1\right)
+\big[\,4\mathcal{H}-2W^2+\left(W^2-w^2\right)\left(1-K_2(q)\right)\big]
\left(1-K_1(q)\right)}}
\,.
\]
Now, choosing $K_1=K_2=K$ and defining the asymptotic speeds $c_1$ and $c_2$ such that
\[
P=c_1+c_2
\,,\qquad\quad
{\cal H}=\frac12\big(c_1^2+c_2^2+w_1^2+w_2^2\big)
\,,
\]
leads to the momentum relation
\[
p^2+w^2=-\frac{4(c_1c_2+w_1w_2)}{1-K}
+(w_1+w_2)^2+(c_1+c_2)^2
\,.
\]
If the smoothing kernel is chosen to satisfy $K(0)=1$ (e.g. inverse Helmholtz or Gaussian), then the quantity $p^2+w^2$ (and thus the momentum $p$, since $w$ is constant) diverges for $q=0$. Defining the vector 
\[
\boldsymbol{\eta}_i=(c_i,w_i)
\]
allows the problem to be classified into two separate cases:

\paragraph{Overtaking collisions:} These satisfy $\be_1\cdot\be_2>0$. Consequently,  the peak separation cannot vanish, since \makebox{$p^2+w^2>0$}. In this case it is possible to express the minimum separation between the peaks by simply setting $p^2=0$:
\[
K\!\left(q_{\rm min}\right)=1-\frac{4(c_1c_2+w_1w_2)}{(c_1+c_2)^2+4w_1w_2}>0
\,.
\]
To visualize such a collision, we take the initial conditions
\[
q_1\left(0\right)=20,\,\, p_1\left(0\right)=-0.4,\,\, w_1\left(0\right)=w_1=1,
\]
\[
q_2\left(0\right)=-20,\,\, p_2\left(0\right)=0.4,\,\, w_2\left(0\right)=w_2=1,
\]
and evolve the two-particle equations~\eqref{eq:ode} forward in time, using
Helmholtz kernels with $\alpha_1=\alpha_2=3$ (these are the values we use
throughout this section).  The
results are shown in Figure \ref{fig:overtaking}.
The particles initially move towards each other, they interact and feel their
mutual repulsion, and then move away from each other indefinitely.  The particles
are always separated from each other by some finite distance and thus they do not change their order.

\paragraph{Head-on collisions:} These possess the property that $\be_1\cdot\be_2<0$.
Thus the peaks can
overlap and the momentum $p$ diverges when this happens. For this particular
example, we choose to study the special case of two completely antisymmetric
peakons, so that $\be_1=-\be_2$ and $q_1=-q_2$. This is an interesting simplification
since it yields $P=Q=W=0$ and the quadrature formula becomes
\begin{equation}
dt=\frac{dK}{\,K^{\,\prime}\sqrt{\,\big[4\mathcal{H}-w^2\left(1-K(q)\right)\!\big]
\!\left(1-K(q)\right)\,}}
\,,
\label{eq:quadrature}
\end{equation}
where
\[
4\mathcal{H}-w^2\left(1-K(q)\right)=p^2\left(1-K(q)\right)\geq0
\,.
\]

Since we are dealing with repulsive interactions, the two particles travel oppositely apart after the collision. Thus, asymptotically $K(q)\to0$, $p\to2c$ and $H\to \left|\be\right|^2=c_0^2+e_0^2$, where $\be=(c_0,e_0)$
(or $\be=-(c_0,e_0)$) collects the asymptotic speed and the charge of
the particle (or antiparticle). Setting $H=c_0^2+e_0^2$ in the equation above
yields the phase trajectory
\begin{equation}
p=\pm\,2\,\sqrt{\,\frac{c_0^2+e_0^2}{1-K(q)}-e_0^2\,}
=\pm\,2\,\sqrt{\,\frac{c_0^2+e_0^{2\,}K(q)}{1-K(q)}\,}
\,,
\label{eq:phase_portrait}
\end{equation}
so that $p$ diverges when $q\to0$, thereby recovering the same qualitative
behavior as the ordinary peakon solution for the CH equation.

\begin{figure}[htb] 
\begin{center}
\subfigure[]{
\includegraphics[width=0.45\textwidth]{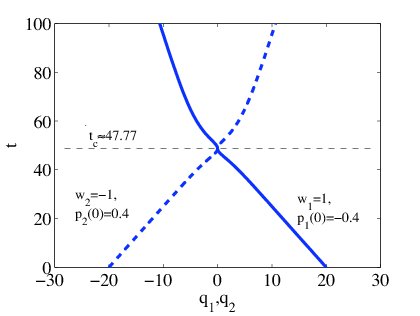}
}
\subfigure[]{
\includegraphics[width=0.45\textwidth]{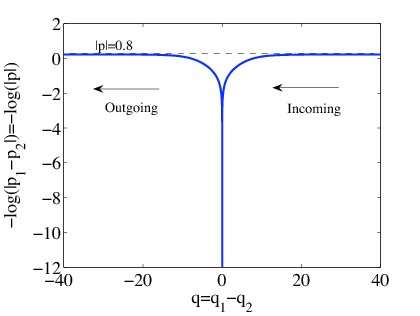}
}
\end{center}
\caption{(a) Particle trajectories for $\be_1\cdot\be_2<0$.
 The difference $q=q_1-q_2$ changes sign at the collision point and, thus, the particle orders are reversed; (b) Phase portrait $p$ versus $q$ for this collision.
 }
\label{fig:headon}
\end{figure}

We obtain a numerical solution in this case by focussing on the intial conditions
\[
q_1\left(0\right)=20,\,\, p_1\left(0\right)=-0.4,\,\, w_1\left(0\right)=w_1=1
\,,
\]
\[
q_2\left(0\right)=-20,\,\, c_0=p_2\left(0\right)=0.4,\,\, -e_0=w_2\left(0\right)=w_2=-1
\,.
\]
The evolution of these initial conditions is displayed in Figure \ref{fig:headon}.
The particles initially move towards each other,
collide, repel each other, and separate.  Notice however that the particle
orders change at the collision point: the particle with label $1$ 
is orignally on the right but after the collision is on the left, and conversely for particle $2$.
 This is in contrast to the case of the overtaking collisions (case 1). 
 A similar order-changing result is observed for the Gaussian kernel. 

Superficially, the idea that the particles exchange order at the collision
time may appear undesirable, and could be rectified by exchanging the particles' momenta after the collision.  This is legitimate because any choice of method for integrating the equations~\eqref{eq:ode} beyond the collision singularity is not unique.  Nevertheless, the order-changing conditions are more natural in this context.  
To see this, we regularize the kernel $K\left(q\right)$
by letting $K\left(q\right)\rightarrow K_{\varepsilon}\left(q\right)=e^{-\left(|q|+\varepsilon\right)/\alpha}$,
so that $K_\varepsilon\left(q\right)$ is strictly less than unity.  
In this case, the momenta are finite at the collision, and the particle trajectories pass through one another.  This indicates that the order-changing conditions on the momenta in the unregularized case are the more natural choice for integration beyond the collision singularity.

We gain some understanding of the mathematical structure of the collision by studying the quadrature formula~\eqref{eq:quadrature}, which in terms of the initial conditions $\left(c_0,e_0\right)$ reads
\[
dt=\frac{dq}{\sqrt{\left[4\left(c_0^2+e_0^2\right)-4e_0^2\left(1-K\left(q\right)\right)\right]\left(1-K\left(q\right)\right)}}
\,,
\]
with solution
\[
t=\phi\left(q\right)-\phi\left(q_0\right),\qquad q\left(0\right)=q_0>0
\,,
\]
where
\begin{multline*}
\phi\left(q\right)=
\frac{\alpha}{c_0}\log\left(c_0^2+e_0^2\right)-\\
\frac{\alpha}{2c_0}\log\left[e^{q/\alpha}\left(2c_0^2-e^{-q/\alpha}c_0^2+e_0^2e^{-q/\alpha}+2c_0\sqrt{-e_0^2e^{2q/\alpha}-e^{-q/\alpha}c_0^2+e_0^2e^{-q/\alpha}+c_0^2}\right)\right].
\end{multline*}
%
%
%
%
The collision time is obtained by setting $q=0$,
\[
t_{\mathrm{c}}=\tfrac{1}{2}\alpha\log\left(c_0^2+e_0^2\right)-\phi(q_0)
\,,
\]
which is finite.  For the starting conditions used in Figure \ref{fig:headon}, we
obtain a collision time $t_{\mathrm{c}}=47.77$, in agreement with the numerical
results shown in the figure.
Finally, Figure \ref{fig:headon}~(b) provides further information about the
collision through a phase portrait; this plot coincides exactly with
formula~\eqref{eq:phase_portrait}.
The momentum is $p=-2c_0=-0.8$ before and after the collision, while during
the collision, it diverges to $p=-\infty$.
%
%
%
%
%
%
%
%
%

\paragraph{Limiting case:} There is also a limiting case when $\be_1\cdot\be_2=0$.
In this case, one finds that $p^2$ assumes the constant value
\begin{subequations}
\begin{equation}
p^2=P^2+W^2-w^2=4w_1w_2+(c_1+c_2)^2
\,.
\end{equation}
Consequently, from the equation
\begin{equation}
\frac{dq}{dt}=p\left(1-K(q)\right)
\label{eq:ode_simple_b}
\end{equation}%
\label{eq:ode_simple}%
\end{subequations}%
we conclude that if $p=-\,\sqrt{P^2+W^2-w^2}$, then the two particles 
merge after the peaks overlap and they never split apart. This happens {\it independently} of the particular choice of $K$.
%
%
%
%
%
%
Other stationary states of the pairwise interaction dynamics may be found
by setting $\dot{q}^2=0$ and by solving in $q$. 

\begin{figure}[htb]
\begin{center}
\includegraphics[width=0.5\textwidth]{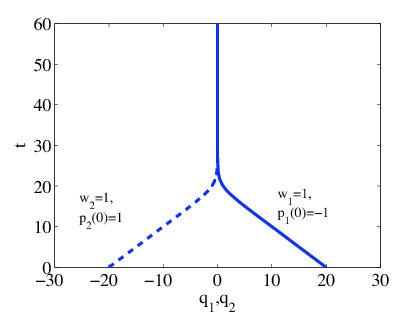}
\end{center}
\caption{Particle separation trajectories for the limiting case $\be_1\cdot\be_2=0$ show that the particles asymptotically approach each other.}
\label{fig:limiting}
\end{figure}

To illustrate the dynamics in this limiting case, we take the following initial
conditions:
\[
q_1\left(0\right)=20,\,\, p_1\left(0\right)=-1,\,\, w_1\left(0\right)=w_1=1,
\]
\[
q_2\left(0\right)=-20,\,\, p_2\left(0\right)=1,\,\, w_2\left(0\right)=w_2=1,
\]
and evolve the two-particle equations~\eqref{eq:ode} forward in time, as shown in Figure \ref{fig:limiting}.

\begin{figure}[h!]
\begin{center}
\includegraphics[width=0.5\textwidth]{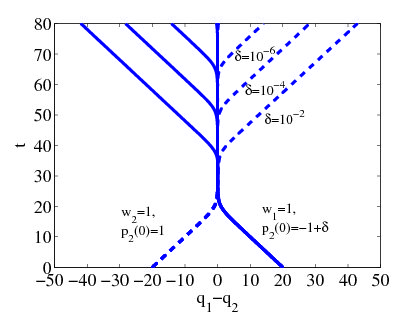}
\end{center}
\caption{
Particle separation trajectories show a phase shift that lengthens as $\be_1\cdot\be_2\rightarrow0$ (through positive values). Apparently the phase shift diverges logarithmically as $\delta\to0^+$.}
\label{fig:limitingsequence}
\end{figure}

As in the attractive case of head-on collisions, the particles initially
move towards each other, and interact.
 In this case, it is easy to understand the mathematical structure of the
 collision because the equations of motion have the simple form exhibited
 by Eqs.~\eqref{eq:ode_simple}, for which $p<0$.  Thus, defining $\tau=|p|t$,
 Eq.~\eqref{eq:ode_simple_b} becomes
\[
\frac{dq}{d\tau}=-\left[1- K\left(q\right)\right].
\]
Using the indefinite integral
\[
\int\frac{dq}{1-K\left(q\right)}=\mathrm{sign}\left(q\right)\alpha\log\left(1-e^{-|q|/\alpha}\right),
\]
%
%
%
%
we obtain the solution
\[
q=-\alpha\log\left(1-C_0e^{-\tau/\alpha}\right),\qquad C_0=1-e^{-q_0/\alpha},\qquad
0<C_0<1
\,.
\]
The collision occurs when $q=0$, that is, when $\tau=\infty$.  (It is clear that the collision times for the Gaussian and top-hat kernels are also infinite in this case.)  Figure \ref{fig:limiting}
thus shows for $\be_1\cdot\be_2=0$ that the particles move asymptotically closer to each other, but never actually reach the collision point. In this case, one may say that the phase shift of the collision diverges. One might ask how the phase shift behaves in the limit as $\be_1\cdot\be_2\rightarrow0^+$ in a sequence of positive values.  Figure \ref{fig:limitingsequence} answers this question by showing that the  particles approach each other and stay together longer and longer in the limit as $\be_1\cdot\be_2\to0^+$. Figure \ref{fig:limitingsequence}  indicates that the phase shift diverges logarithmically in this limit.


\section{Attractive interactions}
\label{attractive-interact}

Evolution by either CH2 or MCH2 yields a geodesic flow on the semidirect product group ${\rm Diff}\,\circledS\,{\cal
F}$ governed by the Lagrangian (\ref{MCH2-Lag}), which is a norm when $g>0$, corresponding to repulsive particles.  
However, one might also consider, for example, gravitational interactions, for which particles with mass undergo attractive interactions. Attractive interactions correspond to $g<0$ in the potential term in $\rhobar$ of the MCH2 Lagrangian (\ref{MCH2-Lag}), which is then no longer a norm. It is known that the change in sign of $g$ preserves integrability of CH2 when $K_2$ is a delta function
\cite{ChLiZh2005,Ku2007}. One may repeat the previous analysis for attractive interactions by considering $g<0$. This is the subject of the present section. First we show numerical results, then we discuss two-peakon interactions.

\rem{
Allowing negative density is another possibility, which represents the presence of both positive and negative charges in the system.
However, we do not consider densities of different sign in this work.
}

\subsection{Numerical results for $g<0$}

\begin{figure}[htb]
\begin{center}
\subfigure[]{
\includegraphics[width=0.45\textwidth]{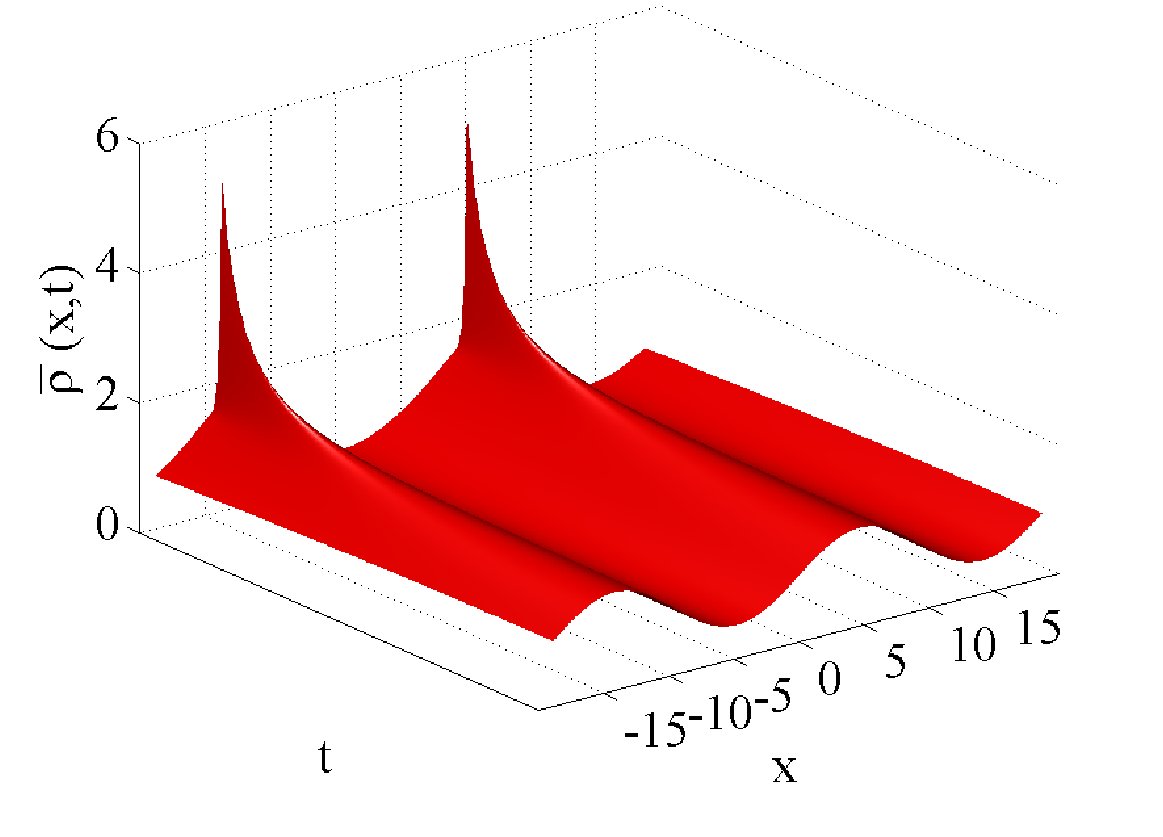}
}
\subfigure[]{
\includegraphics[width=0.45\textwidth]{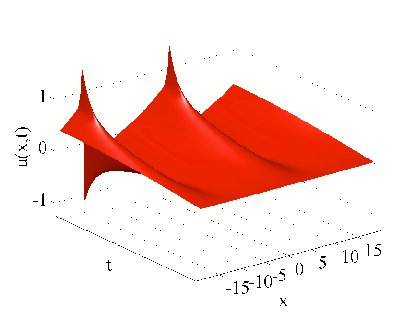}
}
\end{center}
\caption{(Color online) Evolution of the system of equations (\ref{rhobaru-eqns}) for the attractive case $g=-1$.  Subfigure (a) shows the smoothed density $\rhobar $, while (b) plots the velocity $u$.  Singularities in the solutions of both variables emerge after finite time.  The singularities in density $\rho$ form at the maxima of the initial conditions in $\rhobar$. Positive (resp. negative) singularities in velocity $u$ form at the maxima (resp. minima) of the initial conditions in $\rhobar$. 
}
\label{fig:pde_attractive}
\end{figure}

To demonstrate the spontanteous emergence of singular solutions for $g<0$, we numerically solve the equations dealing with attraction.  Specifically, we study equations (\ref{rhobaru-eqns}) with $g=-1$. We
solve these equations on a periodic domain $\left[-L,L\right]$ with the initial conditions
\[
u(x,0)=0.1,\qquad \rhobar\left(x,0\right)=1+\tfrac{1}{2}\sin\left(\frac{4\pi
x}{L}\right)
\,.
\]
This kind of initial conditions was also used in~\cite{HoOnTr2008}
to observe ``clumpons'' in the density $\rhobar $.  We take $\alpha_1=0.3$, $\alpha_2=0.1$.
 The results of the numerical simulations are shown in Figure \ref{fig:pde_attractive}. These results are very similar to those found in \cite{HoOnTr2008}. However, the types of equations involved (Hamiltonian versus gradient flow) are quite different. 
%
%
%
%
%
%

\begin{remark}  
An analytical explanation of the mechanism for the formation of  singularities shown in Figure \ref{fig:pde_attractive} has eluded us, so far. It remains an outstanding problem to prove an ``anti-maximum principle'', that if $\rhobar$ initially has a maximum, then the pointwise $\rho$ must develop a singularity in finite time. Likewise, no theorem exists yet for the formation of the velocity singularities. 
\end{remark}  

\subsection{Pairwise collisions for $g<0$}
The case of attractive potential ($K_2=-K_1$) yields the following momentum relation
\[
p^2
=
-\,
\frac{4(c_1c_2+w_1w_2)}{1-K}
+
\left(W^2-w^2\right)\frac{1+K}{1-K}
+
(c_1+c_2)^2
\,.
\]
Thus, by adding $\left(W^2-w^2\right)/(1-K)$ at each side, one finds
\[
p^2+W^2
=
-\,
\frac{4(c_1c_2-w_1w_2)}{1-K}
+
w^2
+
(c_1+c_2)^2
\,.
\]
Consequently, overtaking collisions are defined by $c_1c_2>w_1w_2$, for which the peaks cannot overlap. The head-on collisions take place when $c_1c_2<w_1w_2$.
The momentum formula above can be obtained from the case of repulsive interactions by exchanging $w_2\to-w_2$. Hence, analysis similar to that in Section \ref{collidingpairs} yields analogous formulas. Also in the limiting case $c_1c_2=w_1w_2$, one recovers the merging phenomenon already observed in the repulsive case. Other stationary states may be found by setting $\dot{q}=0$.
In higher dimensions, these would correspond to equilibria such as the orbits of binary systems typical of  purely gravitational interactions
(\makebox{$K_2=-\,\Delta^{-1}$}).

\rem{ 
Again, for overtaking collision we can evaluate the minimum peak separation
by
\[
K\!\left(q_{\rm min}\right)=1-\frac{4(c_1c_2-w_1w_2)}{(c_1+c_2)^2-4w_1w_2}>0
\]
while head-on collisions with $P=Q=0$ and $w_1w_2=e^2>0$ yield the quadrature formula
\[
dt=\frac{dK_1}{2\,K_1^{\,\prime}\sqrt{
\big[\,\mathcal{H}-e^2\left(1-K(q)\right)\big]
\left(1-K(q)\right)\,}}
\,.
\]
thereby recovering the same qualitative behavior as the case of repulsive
interactions. 

Another kind of head-on collisions is obtained upon considering head-on collisions such that $Q=W=0$ and $c_1c_2=c^2>0$, which yields the following quadrature formula
\[
dt=\frac{dK_1}{2\,K_1^{\,\prime}\sqrt{
\big[\,\mathcal{H}-(c^2-e^2)\left(1+K(q)\right)\big]
\left(1-K(q)\right)}}
\,.
\]

Also in the limiting case $c_1c_2=w_1w_2$, one recovers the same arguments as in the case of repulsive interactions, since the equation $\dot{q}=p(1-K_1)$
is not affected by the sing of $K_2$. Thus, the two peaks merge after overlapping
and they never split out.

}    

\bigskip

\section{Applications of CH2 and MCH2 in imaging science}
\label{metamorph}

Much of the discussion here is motivated by the applications of EPDiff in Computational Anatomy (CA). 
The problem for CA is to determine the minimum distance between two images as specified in a certain representation space, $V$, on which the diffeomorphisms act. Metrics are written so that the optimal path in Diff satisfies an evolution equation. This equation turns out to be EPDiff, when $V$ is a closed contour representing the planar shape of the image. 
The flow generated by the EPDiff equation transforms one shape along a path in the space of smooth invertible maps that takes it optimally into another with respect to the chosen norm. Its application to contours in biomedical imaging, for example, may be used to  quantify growth and measure other changes in shape, such as occurs in a beating heart, by providing the optimal transformative mathematical path between the two shapes. A discussion of EPDiff and the application of its peakons and other singular solutions for matching templates defined by the contours of planar image outlines appears in \cite{HoRaTrYo2004}. 

\subsection*{Metamorphosis}
Metamorphosis is a recent development in the problem of image-comparison for CA. In the metamorphosis of smooth images one considers a manifold, $N$ which is acted upon by a Lie group $G$: $N$ contains what may be regarded as ``deformable objects'' and $G$ is the group of deformations, which is the group of diffeomorphisms in most  applications. Several examples for the space $N$ are treated in \cite{HoTrYo2007}.

\begin{definition}
A {\bfi metamorphosis} \cite{TrYo05} is a pair of curves $(g_t,\,\eta_t)\in G \times N$ parameterized by time $t$,
with $g_0 = \mathrm{id}$. Its {\bfi image} is the curve $n_t\in N$ defined by the action 
$n_t = g_t\eta_t$.
The quantities $g_t$ and $\eta_t$ are called the {\bfi deformation part} of the
metamorphosis, and its {\bfi template part}, respectively. When $\eta_t$ is
constant, the metamorphosis is said to be a {\bfi pure
deformation}. In the general case, the image is a combination of a
deformation and template variation.
\end{definition}

\paragraph{Riemannian metric}
A primary application of the metamorphosis framework is based on the definition of a Riemannian metric on $G\times N$ which is invariant under the action of $G$: $(g, \eta)h = (gh, h^{-1}\eta)$. The corresponding Lagrangian on $TG\times TN$ then takes the form
\[
L(g,\dot{g},\eta,\dot{\eta}) 
= L(\dot{g}g^{-1},g\eta,g\dot{\eta}) 
=: l(u, n, \nu) 
= \|(u, \nu)\|^2_n
\,,
\]
where $u:=\dot{g}g^{-1}\in\mathfrak{g}$ (the Lie algebra $\mathfrak{g}$ of the Lie group $G$), $n:=g\eta$, $\nu:=g\dot{\eta}$ and $\|(u, \nu)\|^2_n$ is a norm on $\mathfrak{g}\times TN/G$ parameterized by $n$. 
The optimal matching problem is now equivalent to the computation of geodesics for the canonical projection of this metric from $G\times N$ onto $N$. 
This construction was introduced in \cite{MiYo01}. 
The interest of this construction is that it provides a Riemannian metric on the image manifold $N$ which incorporates the group actions of $G$.
The evolution equations were derived and studied in \cite{TrYo05}
in the case $l(u, n, \nu) = \|u\|_{\mathfrak g}^2 + \|\nu\|_n^2$, for a
given norm,  $\|\,.\,\|_{\mathfrak g}$, on $\mathfrak{g}$ and a prescribed Riemannian  structure on the manifold $N$. In \cite{HoTrYo2007} the metamorphosis approach to image matching was formulated in terms of Hamilton's principle in the Euler-Poincar\'e variational framework \cite{HoMaRa1998} to derive the evolution equations.  This formulation provides an interesting contrast between the variational formulations of optimal control problems and evolutionary equations. 

\paragraph{Semidirect product Lie groups}
Assume that $N$ is a Lie group and that for all $g\in G$, the action of
$g$ on $N$ is a group homomorphism. That is, for all $n, \tilde n\in N$,
$g(n\tilde n) = (gn)(g\tilde n)$. For example, $N$ can be a vector space and the action of $G$ can be linear. Consider the semidirect
product $G \circledS N$ with  $(g,n)(\tilde g ,\tilde n)  =(g\tilde g,
(g\tilde n)n)$ and build on $G\circledS N$ a right-invariant
metric constrained by its value $\|\,.\, \|_{(\id_G,\id_N)}$ at the
identity. Then optimizing the geodesic energy in
$G\circledS N$ between $(\id_G, n_0)$ and $(g_1, n_1)$ with fixed $n_0$ and $n_1$ and free $g_1$ yields a particular case of metamorphosis.

The Euler-Poincar\'e formulation of metamorphosis on ${\rm Diff}\,\circledS\,G$ was also considered in \cite{HoTrYo2007}. The Euler-Poincar\'e equations for a Lagrangian $l(u,\nu)$ in which the variable $n$ is absent are found for $G\circledS N={\rm Diff}\circledS \mathcal{F}$ to produce the CH2  and MCH2 systems for the corresponding norms,
\[
l_{CH2}(u,\nu)=\|u\|_{H^1}^2 + \|\nu\|_{L^2}^2
\quad\hbox{and}\quad
l_{MCH2}(u,\nu)=\|u\|_{H^1}^2 + \|\nu\|_{H^1}^2
\,.
\] 
These are precisely the norms discussed in the present paper.
Thus, the theory of metamorphosis in imaging science summons the singular solutions for MCH2, when the smoothing kernels $K_1$ and $K_2$ are chosen to be Helmholtz inversions. Future developments will determine the utility of these singular solutions in  imaging science. 
\smallskip

Actually, the problem of matching shapes in imaging science is an optimal control problem, rather than the initial value problem (IVP) discussed here. However, as we have seen, the IVP for MCH2 is illuminating and interesting in its own regard. Moreover, considerations of the IVP help in the interpretation of the solutions of the optimal control problem for image matching as a flow of information parcels (landmarks) from one image to another. The particle-like solution behavior for the IVP discussed here is expected to hold for any symmetric, confined, translation-invariant choices of the smoothing kernels $K_1$ and $K_2$.


\bigskip
\subsection*{Acknowledgements}
It is a pleasure to thank C. J. Cotter, J. D. Gibbon, R. Ivanov, V. Putkaradze and L. Younes for many illuminating discussions of these matters. The work of Holm and Tronci was partially supported by a Wolfson Award from the Royal Society of London. We are also grateful to the European Science Foundation for partial support through the MISGAM program.

\bigskip

\bibliographystyle{unsrt}

\end{document}